\documentclass[prl,twocolumn,showpacs,amsmath,amssymb,floatfix,superscriptaddress]{revtex4}

\usepackage[]{graphicx}         	
\usepackage{bm}             		
\usepackage{tabularx}
\usepackage{times}				

\newcommand{\IEF}{Institut d'Electronique Fondamentale, CNRS, UMR 8622, 91405 Orsay, France}
\newcommand{\UPS}{Universit{\'e} Paris-Sud, 91405 Orsay, France}
\newcommand{\IMEC}{IMEC, Kapeldreef 75, B-3001 Leuven, Belgium}
\newcommand{\USFD}{Department of Engineering Materials, University of Sheffield, Sheffield S1 4DU, United Kingdom}

\begin{document}

\title{Auto-oscillation threshold and line narrowing in MgO-based spin-torque oscillators}

\author{S. Cornelissen}
\affiliation{\IMEC}
\author{L. Bianchini}
\affiliation{\IEF}
\affiliation{\UPS}
\author{G. Hrkac}
\affiliation{\USFD}
\author{M. {Op de Beeck}}
\author{L. Lagae}
\affiliation{\IMEC}
\author{Joo-Von Kim}
\email{joo-von.kim@u-psud.fr}
\author{T. Devolder}
\author{P. Crozat}
\author{C. Chappert}
\affiliation{\IEF}
\affiliation{\UPS}
\author{T. Schrefl}
\affiliation{\USFD}

\date{\today}                                           

\begin{abstract}
We present an experimental study of the power spectrum of current-driven magnetization oscillations in MgO tunnel junctions under low bias. We find the existence of narrow spectral lines, down to 8 MHz in width at a frequency of 10.7 GHz, for small applied fields with clear evidence of an auto-oscillation threshold. Micromagnetics simulations indicate that the excited mode corresponds to an edge mode of the synthetic antiferromagnet.
\end{abstract}

\pacs{75.75.+a, 72.25.Pn, 85.75.-d}

\maketitle

Spin-transfer torques~\cite{Slonczewski:JMMM:1996,Berger:PRB:1996}, which involve the transfer of spin-angular momentum between spin-polarized currents and magnetization in magnetic multilayers, lead to many interesting possibilities for controlling magnetization dynamics in nanoscale devices. One important application concerns nanoscale radiofrequency (rf) oscillators, in which magnetization oscillations, driven by a DC spin-polarized current, are tunable with applied field and current with the potential for large power output. Such current-induced magnetization oscillations have been observed in metallic spin-valve nanopillars~\cite{Kiselev:Nature:2003} and nanocontacts~\cite{Rippard:PRL:2004}. However, there is only a limited body of work~\cite{Nazarov:APL:2006,Petit:PRL:2007,Houssameddine:APL:2008,Deac:NP:2008} to date concerning clear spin-torque oscillations in magnetic tunnel junctions (MTJ) with large tunnelling magnetoresistance ratios (TMR), particularly for MgO-based junctions. An experimental realization of large-amplitude oscillations in such MTJ systems, which would allow sufficiently large power to be generated for applications, is therefore an important scientific problem.
 
In this article we present experimental evidence of magnetization auto-oscillations in MgO-based tunnel junctions which are in the ``virgin'' state, i.e. before noticeable deterioration of the tunneling characteristics due to large sustained applied voltages across the junction. We show that a clear auto-oscillation threshold is reached for increasing current, which is accompanied by a drastic increase of the emitted power and line narrowing. Furthermore, narrow spectral lines, with linewidths on the order of a few MHz for oscillation frequencies in the range of 8-15 GHz, are observed even at \emph{zero} applied fields. This is remarkable because high-frequency power spectra with comparable linewidths have only been observed in nanopillars until now in spin-valve systems under large applied fields~\cite{Kiselev:Nature:2003,Mistral:APL:2006}. As we will show, micromagnetics simulations indicate that the observed mode is an edge mode of the synthetic antiferromagnet.

The measurements were performed on low resistance-area (RA) product MgO-based magnetic tunnel junctions with layer composition of Ta (3)/CuN (40)/Ta (5)/PtMn (20)/ CoFe$_{30}$(2) /Ru(0.8)/CoFe$_{20}$B$_{20}$(2)/Mg (1.3 [nat ox])/CoFe$_{20}$B$_{20}$(3)/Ta(8) (thicknesses in nm) deposited on a Singulus Timaris tool. These junctions were patterned using electron-beam lithography and ion beam etching into rectangular nanopillars with a nominal size of 100$\times$200 nm$^2$. The electrical resistance of the tunnel junctions in the ``virgin'' state are typically 180 and 280 $\Omega$ for the parallel and antiparallel configurations, respectively, which correspond to a TMR ratio of 55\%. The RA product is 0.9 $\Omega \, \mu\textrm{m}^2$. Similar results were obtained with samples having an RA product of 1.4 $\Omega \, \mu\textrm{m}^2$. The saturation magnetization $\mu_0 M_s$ is deduced to be 1.2 T in the patterned samples, while FMR measurements on unpatterned layer stacks show the effective magnetization to be 1.9 T, the Land{\'e} factor to be 2.14 and the uniaxial anisotropy to be 2.5 mT. Details concerning the characterization of the magnetic properties are presented elsewhere~\cite{Cornelissen:JAP:2009}.

For the electrical measurements, the applied current through the tunnel junction is generated by applying a DC voltage to a 50 ohm resistor in series with the sample. Two bias tees allow DC and RF routing. One of the RF routes is amplified by 25 dB in a frequency band of 100 MHz $-$ 26 GHz and connected to a spectrum analyzer. The resolution bandwidth of the analyzer is set to 5 MHz for scans across the entire 0.1-26 GHz range, while zoomed scans are performed with a resolution bandwidth of 1 MHz. The final power spectra are obtained after subtracting a reference curve taken with a low current bias. For positive currents, electrons flow from the SAF system to the free layer, which favors a parallel magnetization state in experiment. Material fatigue occurs for biases above 0.4 V. In the study presented here, we use low biases from $-1.6$ to $+1.6$ mA to study power spectrum in ÒvirginÓ state devices. The rf measurements are made in the presence of applied magnetic fields in the film plane, which are generated with electrical coils or an electromagnet. In all our measurements a positive magnetic field along the easy axis favors the parallel magnetization configuration of the magnetic tunnel junction.

An example of a clear threshold behavior is presented in Fig. 1, where the power spectrum is shown for different applied currents below and above the auto-oscillation threshold.
%
\begin{figure}
\includegraphics[width=8.5cm]{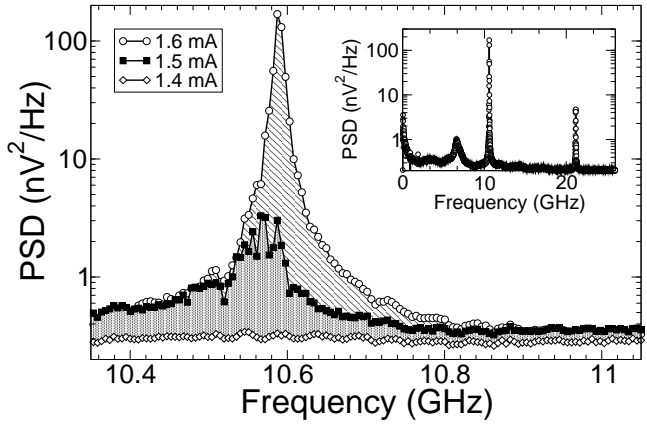}
\caption{\label{fig:threshold}Evidence of auto-oscillation threshold. Power spectral density (PSD) of the voltage noise for applied currents from 1.4 mA to 1.6 mA. Inset: Power spectrum for $I=1.6$ mA over the entire measurement bandwidth of 0.1-26 GHz.}
\end{figure}
The current threshold is estimated to be around $I = 1.5$ mA. When applying $I=1.5$ mA through the device a peak appears at 10.59 GHz in the power spectrum that is well above the noise floor, in contrast to a relatively flat spectrum for $I \leq 1.5$ mA. If the current is increased further above the threshold value, a large increase in the total power is measured and the linewidth at half maximum reduces to a few MHz, with 8 MHz being the narrowest linewidth observed at room temperature and under the experimental conditions described. These spectra were measured at \emph{zero} applied magnetic fields but after preparing the magnetic tunnel junction in the antiparallel (AP) configuration, i.e. between every measurement we applied a strong field to saturate the free layer magnetization along the negative easy-axis direction. The frequency of the peak is not observed to vary much with the current above threshold, in contrast to free-layer excitations above threshold for which a large redshift is expected for in-plane fields. The influence of the current can also be seen in the Stoner-Wohlfarth astroid for in-plane fields as a strong distortion occurs under positive current with respect to a low current astroid. The positive current leads to an additional torque that favors the parallel (P) state. At very low frequencies, a large contribution from $1/f$ noise is observed. At low current thermal modes are also visible (not shown). From the variation of the frequency of the thermal FMR mode as a function of applied field, we determined the effective saturation magnetization $M_{\rm eff}$ from fits to the Kittel formula, $\omega = \gamma \sqrt{(H_a + H_k)(H_a + H_k + M_{\rm eff})}$, where $H_a$ is the applied field and $H_k$ is the anisotropy field, from which we find $M_{\rm eff} = 1.2$ T.

\begin{figure}
\includegraphics[width=8.5cm]{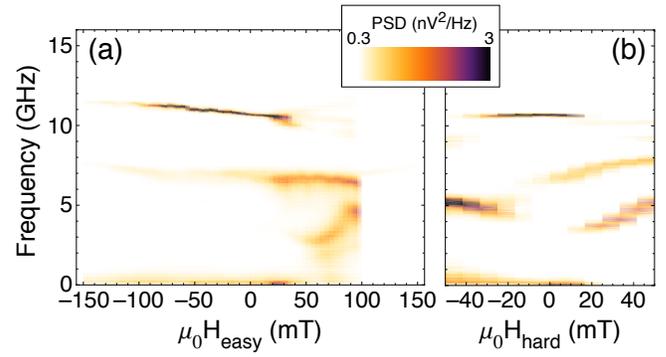}
\caption{\label{fig:map}(Color online) Density plot of the power spectrum as a function of applied magnetic field along the (a) easy axis and (b) hard axis, for an applied current of $I = $ 1.6 mA.}
\end{figure}

The full measured power spectra across the range of 0.1-16 GHz, as a function of applied field along the easy and hard axes, are presented in Fig.~\ref{fig:map} as a color map. 
%
In scans along both field directions, we observe a number of peaks in the power spectrum but only one peak attains large power ($>$ 100 nV$^2$/Hz), which corresponds to the peak shown in Fig.~\ref{fig:threshold}. A second harmonic of this peak is also observed (as shown in the inset of Fig.~\ref{fig:threshold}), around 20-22 GHz, but this is not presented in the color map for the sake of clarity. Third harmonics were not seen as the frequencies are beyond the scope of the apparatus. It is interesting to note that the main peak persists and retains a large power even at zero applied field, which can easily be seen for field sweeps along both the easy and hard axis directions. Moreover, the intensity of the peak is observed to increase as the applied field magnitude is decreased. Other peaks corresponding to thermally-activated modes, in the range of 5-7 GHz, are present over a wide range of the fields studied. A more detailled discussion of these modes will be presented elsewhere.

\begin{figure}
\includegraphics[width=8.5cm]{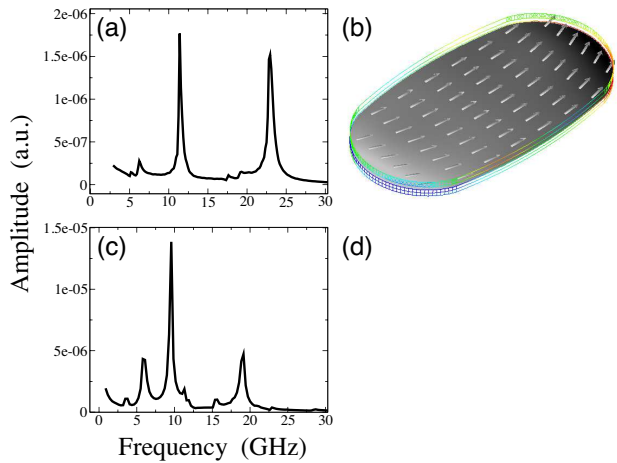}
\caption{\label{fig:sim}Simulated power spectral density of edge modes in the synthetic antiferromagnet, for an applied field of 5 mT along the (a) easy and (c) hard axes. The micromagnetic configuration of the top synthetic antiferromagnet layer (SAF1), illustrating the excited edge mode, are shown in (b) and (d) for the easy- and hard-axis fields, respectively. Good quantitative agreement is found between the simulated and experimental frequencies of the main excitation peak.}
\end{figure}

Micromagnetic simulations show that the observed modes correspond to \emph{edge} modes of the synthetic antiferromagnet bilayer. The simulations consist of tracking the time evolution of magnetization of the full free layer/synthetic antiferromagnet/antiferromagnet system using the same materials, film thicknesses and TMR ratio as the experimental system. In our notation for the synthetic antiferromagnet, SAF1 refers to the magnetic layer adjacent to the free layer and SAF2 refers to the magnetic layer pinned by the PtMn antiferromagnet. The system is discretized with a finite element method with a linear basis function and a discretization size of 2 nm, which is below the exchange length of 4.5 nm~\cite{Ertl:JAP:2006}. Three different calculations were performed for a given applied current and applied field geometry. In the first, a mutual spin-torque is applied to all magnetic layers self-consistently. In the second, the spin-torque is applied only on the free layer, and in the third, the spin-torque is applied only on SAF1. This allows us to clearly identify the excitations in the spin-torque--coupled free layer/SAF1 layer system, the free layer, and the SAF1 layer separately. The threshold current for oscillations in the simulations, 6 mA, is higher than the observed experimental value of 1.5 mA, which can be explained by differences in spin polarization or other transport parameters. For an applied field of 5 mT along the easy direction we observe SAF1 layer excitations at 11.5 GHz and 23.1 GHz (second harmonic), as shown in Figure~\ref{fig:sim}.
%
In a second simulation run we applied an external field of 5 mT in the hard axis direction and observed a redshift in the frequency, in contrast to the behavior for easy axis field variations and in agreement with the experimental observations. For applied fields along both the easy and hard axis directions we find good quantitative agreement with the experimental frequencies.

\begin{figure}
\includegraphics[width=8.5cm]{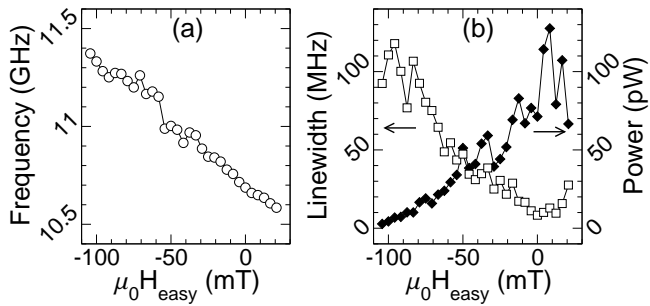}
\caption{\label{fig:fits}(a) Frequency, (b) linewidth (squares) and integrated power (diamonds) of the main peak as a function of applied field along the easy axis, with $I=$ 1.6 mA.}
\end{figure}

The variation of the frequency, linewidth, and total power of the experimental peak, as a function of the easy axis field under an applied current of 1.6 mA, is presented in Fig.~\ref{fig:fits}.
%
These parameters were extracted using a Lorentzian fit to the power spectra. As we mentioned above, the linewidth decreases as the applied field is reduced. Peaks observed in this low-field region are much narrower than those at higher fields, with a minimum in the linewidth of 8 MHz observed at \emph{zero} applied field, after which the line broadens again as the switching transition is approached. Such narrow lines are consistent with the auto-oscillation regime~\cite{Kim:PRL:2008,Kim:PRL:2008b}. Some jitter in the spectral lines was observed by measuring the spectra at different sweep frequencies, which indicates that inhomogeneous broadening may be an important factor for narrow lines and that the actual oscillator coherence is much better than what the measured linewidths suggest. Further experimental studies on this issue are currently in progress. The maximum total power is observed at low applied field, which indicates that the oscillation, in addition to better coherence, also gains in amplitude. A similar behavior for the linewidth and power was observed for field sweeps along the hard axis, with the exception of a much smaller window of existence for the excited mode, which can easily be ascertained from Fig.~\ref{fig:map}b.

In summary, we provide clear experimental evidence of current-driven magnetization oscillations above the auto-oscillation threshold in magnetic tunnel junctions based on MgO barriers. We have brought to light the possibility of obtaining narrow spectral lines, down to 8 MHz at a frequency of 10.7 GHz, at zero applied magnetic fields, in tunnel junctions close to their ``virgin'' state. Micromagnetics simulations show that the excited mode is the edge mode of the synthetic antiferromagnet bilayer.

We thank Dr. W. Maass (Singulus Technologies AG) for supplying the multilayers. This work was supported by the European Communities programs IST STREP, under Contract No. IST-016939 TUNAMOS, and ``Structuring the ERA'', under Contract No. MRTN-CT-2006-035327 SPINSWITCH, and by the local government of R{\'e}gion Ile-de-France within the framework of C'Nano IdF. SC acknowledges IWT Flanders for financial support.

\bibliography{articles}

\end{document}